\documentclass[twocolumn,prl,aps,epsf,showpacs]{revtex4}

\usepackage{amssymb,epsf}
\usepackage{graphicx}
\usepackage{epstopdf}

\begin{document}

\title{Competition Between Fractional Quantum Hall Liquid, Bubble and Wigner Crystal
  Phases in the Third Landau Level}

\author{G. Gervais$^{1,2}$, L.W.  Engel$^{2}$,  H.L. Stormer$^{1,3}$, D.C. Tsui $^{4}$, K. W. Baldwin$^{3}$, K.W. West$^{3}$, and L.N. Pfeiffer$^{3}$}

\address{$^{1}$Department of Physics and Department of Applied Physics, Columbia University, New York, NY 10027\\
$^{2}$National High Magnetic Field Laboratory, Tallahassee, FL 32306, USA\\
$^{3}$Bell Laboratories, Lucent Technology, Murray Hill, NJ 07974\\
$^{4}$Department of Electrical Engineering, Princeton University
Princeton, NJ 08544\\}

\begin{abstract} 
Magnetotransport measurements were performed in a ultra-high mobility 
GaAs/AlGaAs quantum well of density $\sim 3.0 \times 10^{11}$ $cm^{-2}$. 
The temperature dependence 
of the magnetoresistance $R_{xx}$ was studied in detail in the vicinity of $\nu=\frac{9}{2}$.
In particular, we discovered new minima in $R_{xx}$ at filling factor $\nu\simeq 4\frac{1}{5}$ and $4\frac{4}{5}$, but
only at intermediate temperatures $80\lesssim T\lesssim 120$ mK. We interpret these 
as evidence for a fractional
quantum Hall liquid forming in the N=2 Landau level and competing with
bubble and Wigner crystal phases favored at lower temperatures. Our data
suggest that a magnetically driven insulator-insulator  quantum phase transition occurs between the bubble and Wigner crystal phases at $T=0$.

\end{abstract}
\pacs{73.43.-f, 73.43.Np,73.63.Hs} 

\maketitle
\narrowtext

The fractional quantum Hall effect (FQHE) in a two-dimensional electron gas (2DEG) is
 typically understood  as an incompressible Laughlin liquid of electrons giving 
  rise at low temperatures to a vanishing magnetoresistance, $R_{xx}$ concomitant with  
a  quantized Hall resistance, $R_{xy}=h/ \nu e^{2}$, where $\nu$ is the Landau level (LL) filling
factor. This effect is now widely described in terms of the composite fermion model (CF)
 \cite{Jain89a,Jain89b,Lopez91,Halperin93}, and  has been so far observed only in the first (N=0) and second (N=1) Landau levels. This experimental fact is in agreement with early 
calculations based on exact diagonalization of a small particle ensemble, which suggest
that a Laughlin (or FQHE)  liquid is energetically unfavored in 
 the higher Landau levels, $N>1$\cite{Ambrumenil88,Belkhir95,Morf95}. In this Letter, we
report on new magnetotransport measurements performed in a ultra-high mobility 
 GaAs/AlGaAs specimen which show evidence, at
intermediate temperatures $80\lesssim T\lesssim 120$ mK, for a FQHE liquid forming
in the N=2 Landau level at filling factor $\nu\simeq4\frac{1}{5}$ and $4\frac{4}{5}$.

Previous measurements performed
in ultra-high mobility heterostructures have shown, in the lowest  Landau level, the CF model to 
correctly account for the sequence of FQHE of electrons in terms of an integer quantum Hall effect (IQHE) of CF particles.
Recent, newly discovered FQHE states  such as, for example,  at $\nu=\frac{4}{11}$ or the $\nu=\frac{7}{11}$, find in the CF model an elegant interpretation in terms of a FQHE of these flux-attached 
CF particles\cite{Pan03}. While all FQHE states
in the lowest Landau levels are well understood within such a model, the current situation
differs much even in the second (N=1) Landau level. For example, insulating
and re-entrant integer quantized Hall effects (RIQHE) have recently been reported in the vicinity
of $\nu=\frac{5}{2}$ and $\frac{7}{2}$ \cite{Eisenstein01,Xia04}.

The phenomenology observed in higher LL $(N>1)$
is even more complex. Several measurements in these LL have found  new and distinct phenomena 
 from those observed in lower LLs. For example, a strong resistance
anisotropy is observed at half-integer filling $\nu=\frac{9}{2}$, together with insulating phases existing in  its flank\cite{Lilly99,Du99}, giving rise to a RIQHE\cite{Cooper99} around filling
factor $\nu\simeq 4\frac{1}{4}$ and $4\frac{3}{4}$. These experimental observations certainly point
toward  many-body correlations of a different lineage than at $\nu<1$, and  
in particular are interpreted as `stripe' and `bubble' charge density waves\cite{Fogler96,Koulakov96,Moessner96,Rezayi99,Haldane00}  as well as liquid crystalline 
phases \cite{Fradkin99}.  The question on whether or not a FQHE liquid can form in the $N>1$ Landau levels has been reexamined  by Fogler and Koulakov\cite{Fogler97}.
Experimental insights on the existence, or non-existence, of a Laughlin liquid in the $N>1$ LL is therefore
timely and highly relevant to our understanding of electron-electron correlations in these higher LLs.

The magnetotransport measurements were performed on a 300 $\AA$  wide modulation-doped 
GaAs/AlGaAs quantum well, of density $n\simeq 3.0 \times 10^{11}$ $cm^{-2}$ and ultra-high
mobility $\mu \simeq 27\times 10^{6}$ $cm^{2}/Vs$. Electrical contact to the 2DEG was achieved via diffusion 
of small indium beads into the semiconductor surface. The electronic transport was measured at
a fixed current varying from  4 to 10 $nA$, and using a quasi-dc lock-in technique (2-10 Hz). 
The sample was mounted in
the  $^{3}He-^{4}He$ mixture of a top-loading dilution refrigerator. The temperature
reported in this work is that of the helium mixture in the mixing chamber, measured with a calibrated ruthenium
oxide thermometer;  at temperature $T\gtrsim 30$ mK, the electron bath
is  assumed to be in good thermal contact with the thermometer. The overall accuracy of the thermometry 
at temperatures above 30 mK (including the correction for the magnetoresistance) is estimated to be within 
10$\%$ of the value reported.

\begin{figure}[t]
\begin{center}
\includegraphics[width=8cm]{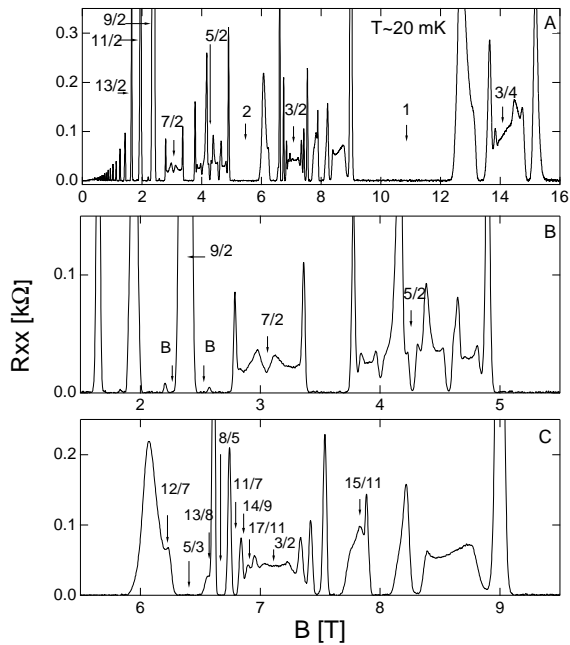}
\end{center}
\vspace*{-5mm}\caption{  {\small Longitudinal resistance $R_{xx}$ as a function of magnetic field measured
with an excitation current of 4 nA. In the panel A, $R_{xx}$ is shown over the whole field range measured between
0 and 16T, with the principal filling factor denoted with arrows. Panels B and C show the same data, but
rescaled such as to show the smaller features. New developing minima in panel C are labeled 
with their tentative fractional assignment, $\frac{12}{7}$,  $\frac{13}{8}$ and $\frac{15}{11}$. The RIQHE state near $\nu=\frac{9}{2}$ is labeled with the letter B.}}
\label{fig1}
\end{figure}

Before turning to the main subject of the paper, we first describe some salient features observed
 in this exceptionally high-quality sample. The longitudinal resistance $R_{xx}$ at temperature $T\sim20$ mK is shown
in Figure 1, panel A, over the magnetic field range spanning from 0 to 16T. Enlarged plots
for various field ranges are shown in panel B and C to resolve the finer structures.
The electronic density was determined from known well-developed  FQHE states to be 
 $n=2.58 \times10^{11}$ $cm^{-2}$. We added, as a guide, several principal filling factors and have shown them as arrows.
The minima in $R_{xx}$ near $\nu=\frac{9}{2}$ associated with RIQHE  are labeled with the letter B. 
In terms of the anisotropic $\nu=\frac{9}{2}$
terminology, these data were taken in a contact configuration in the hard direction, i.e. with
a maximum in resistance flanking $\nu=\frac{9}{2}$. The exceptional high-quality of
 the sample can be readily seen near $\nu=\frac{5}{2},\frac{3}{2}$ where  several
 minima in $R_{xx}$  which are typically  observed only in the very best 
 samples are well developed \cite{Pan99}. 
 
 The series of fractional states observed
 around $\nu=\frac{3}{2}$ is very well accounted for in the CF model which
 interpret these as series of FQHE states emanating from $\nu=\frac{3}{2}$ and
 given by $\nu=(3p\pm 2)/(2p \pm 1)$, where $p$ is 
 the CF filling factor.  The filling factors $\frac{5}{3},\frac{8}{5},\frac{11}{7},\frac{14}{9}$ and  $\frac{17}{11}$ corresponding to the CF serie (+) with  $p=1,2,3,4$ and $5$ are shown
 as arrows in the panel C of Fig.1.  In addition to the known states, 
  we also observe three new developing minima in $R_{xx}$ at filling factors 
  $\nu\simeq1.72$,  $\nu\simeq1.62$,  and $\nu\simeq1.35$, and labeled in Fig.1
with  their proposed tentative fractional assignment,  $\frac{12}{7}$, $\frac{13}{8}$ and $\frac{15}{11}$, respectively.
 The small discrepancy between the observed minima and the proposed 
 fractional assignment is believed to be either from the shallowness
 of these new minima, or from spin effects known to occur in the first excited LL, and understood in terms of  a CF  model in the presence of a spin \cite{Du95}. These new minima cannot be  explained by the CF series mentioned above. However, the prospect of observing an  FQHE 
 at $\frac{15}{11}$ is particularly  interesting, for it could
 be interpreted, similarly to the $ \frac{7}{11}$, as a FQHE of CF's \cite{Pan03}, but in the upper spin branch of the first LL.

 \begin{figure}[t]
\begin{center}
\includegraphics[width=8.cm]{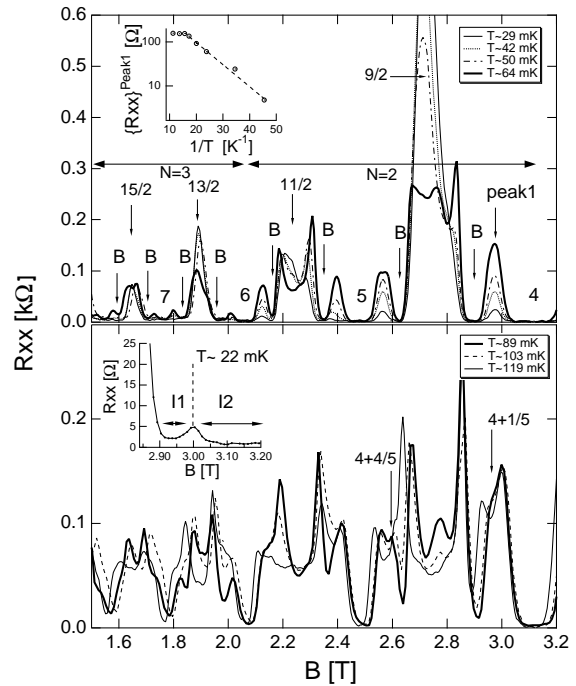}
\end{center}
\vspace*{-5mm}\caption{  {\small  Magnetotransport $R_{xx}$ at various temperatures. The upper (lower)
panel shows the data taken below(above)  $T\lesssim65$ mK.
The half-integer filling factors are indicated with arrows. 
The position of the $\nu=4\frac{1}{5}$
and $\nu=4\frac{4}{5}$ fillings are indicated in the lower panel, together with the bubble
states in the upper panel, labeled with the letter B. The horizontal arrows indicate the 
magnetic field range spanned by the N=2,3 Landau levels. 
The upper inset shows an Arrhenius plot of the
temperature dependence of the resistance maximum labeled peak1. The inset in the lower
panel shows an enlarged version around $\nu=4\frac{1}{5}$ at $T\sim 22 $ mK, with I1 and I2
denoting the two insulating phase. }}
\label{fig3}
\end{figure}

 Figure 2 summarizes our most remarkable findings in the N=2,3 LL.  The magnetotransport
 between $\nu=\frac{9}{2}$ and $\nu=\frac{15}{2}$ was studied in detail at various temperatures.
The higher electron
 density $n=3.02\times 10^{11}$ $cm^{-2}$   than in the data of Fig.1  results from  different illumination
 conditions with a light emitting diode (LED)  during the cooldown of the sample to low temperatures.
 Of particular interest in the upper panel is the existence of minima, B, 
in the flank of the half-integer filling factors $\nu=\frac{9}{2},\frac{11}{2},\frac{13}{2}$ and $\frac{15}{2}$.
These minima are located approximately at filling factor $\nu^{\ast}\equiv\nu-\left[\nu\right] \simeq\frac{1}{4}$ and $\frac{3}{4}$,  where $\left[\nu \right] $ is defined as the largest integer smaller than $\nu$, and have been previously observed 
in DC transport by Lilly {\it et al.}\cite{Lilly99} and Du {\it et al.}\cite{Du99}, and more recently as a microwave resonance by Lewis {\it et al.}\cite{Lewis02}. 
These RIQHE  were interpreted in terms of a triangular crystal lattice phase with 
at least two electrons per lattice site, the so-called `bubble' phase \cite{Fogler96}. 
The data in the upper panel of Fig.2 represents a particularly good case of  the  `bubble'  phase 
persisting up to  the N=3 LL.   

The temperature dependence in the vicinity of  $\nu=\frac{9}{2}$ is particularly  interesting, 
for we observe a crossover from a maximum (peak) to  a minimum (dip) of resistance, 
at $T\sim 65$ mK. A similar crossover is also observed in the temperature dependence of the
 resistance peak separating the bubble state at $\nu^{\ast}\sim\frac{1}{4},\frac{3}{4}$
 and the IQHE at $\nu=4,5$, respectively. For these resistance maxima,  
 a strong temperature dependence is observed at $T\lesssim 65$ mK, while for greater temperatures 
the resistance peak remains constant and new structures are forming in the vicinity of the 
peaks, which we discuss below in details. The 
 temperature dependence of the resistance peak labeled `peak1' is explicitely given in
 the inset of the upper panel of Fig.1 in an Arrhenius plot. The dotted line is a fit
 to the lower temperature data which show an activated behaviour $R_{xx}^{peak1}\propto e^{-E/T}$
 in the range  $15\lesssim1/T\lesssim $50 $K^{-1}$, with a characteristic 
 energy scale $E\sim 0.11(1)$ K. At temperature
 above $\sim$65 mK,  (or $1/T\sim 15$ $K^{-1}$) the saturation of the resistance
 signals the onset of a different regime.
 The inset in the lower panel of Fig.2 shows an enlarged plot of the resistance
 maximum separating the bubble phase\cite{Cooper99,Lewis02} from the IQHE, at the lowest temperature reached in this experiment, 
 $T\sim$ 22 mK.  A small residual resistance peak, of amplitude 
 $\sim5$ Ohm separates the two insulating phases, labeled
 I1 and I2. Previous micro-wave conductivity measurements
 found distinct  resonances in both insulating phases, and interpreted
 these in terms of bubble (I1) and Wigner crystal (I2) phases, respectively \cite{Lewis02,Chen03,Lewis04}. The data in Fig.2 suggest
 that $R_{xx}^{peak}\rightarrow 0$ as $T\rightarrow 0$, 
opening up the interesting possibility of a magnetically driven insulator-insulator  
quantum phase transition between bubble and Wigner crystal phases at $T=0$. 

Recent microwave
measurements in the same filling factor region as the position of the resistance peaks separating
bubble and Wigner crystal phases found evidence for a coexistence region
between the two insulating phases at $T\sim 35$ mK\cite{Lewis04}. On the basis of these
measurements, we interpret our finite resistance peak 
near $\nu^{\ast}\simeq\frac{1}{5},\frac{4}{5}$ as
originating from poorly localized electronic states forming in this mixed insulator phase region.
In the limit $T\rightarrow 0$, 
all electrons in the uppermost Landau levels are fully localized, and ordered  into
either Wigner (I2)  or bubble (I1) crystal phases. As the temperature is increased, 
conduction channels open up as the electrons at the boundary between
I1 and I2 delocalize.  These electronic excitations which show an activated behaviour 
contribute to the diagonal component of the conductivity tensor, 
$\sigma_{xx}=\rho_{xx}/(\rho_{xx}^{2}+\rho_{xy}^{2})$, thus giving rise to 
the non-zero resistance observed, $R_{xx}\propto \rho_{xx}$. The change
in temperature dependence of  $R_{xx}^{peak}$  
together with the appearance of new structures/minima 
located in the peaks, suggest that these unlocalized 
electronic states may give rise to a FQHE at temperatures $T\gtrsim 65$ mK.

The observation of new  structures/minima in  $R_{xx}$ at filling factor near 
$\nu=4\frac{1}{5}$ and $\nu=4\frac{4}{5}$  is striking, for it suggests that
an incompressible Laughlin liquid may be able to form in the N=2 LL.
These two new structures, shown in the lower panel of Fig.2, reside
on the resistance peaks separating bubble and Wigner crystal phases, and
are only observed at intermediate temperatures,  $80\lesssim T\lesssim 120$ mK. 
We regard these minima as the first experimental evidence of a
FQHE liquid forming in the third Landau level.

The disappearance of the minima at lower temperatures can
be interpreted as a competition with I1 and I2  below  $T\sim 65$ mK, in a manner
reminiscent of the FQHE states observed at very small filling, $\nu<\frac{1}{5}$, and obeying
similar energetics\cite{Pan02}. Combining our observations with that
of Lewis {\it et al.}\cite{Lewis04} suggest the following phase diagram
in the third Landau level (lower spin branch). At finite temperatures in the 
filling factors region encompassing the resistance peaks
observed at $4\frac{1}{5}$ and  $4\frac{4}{5}$, there is simultaneous 
coexistence of bubble and Wigner crystal phases, together with 
thermally activated delocalized electronic states present in the same region.  
At sufficiently high temperatures, an excited fractional quantum Hall liquid 
forms in this coexistence phase, but yields to the 
electron solid phases at lower temperatures as the
number of  delocalized states decreases. The resistance peaks observed 
at  $4\frac{1}{5}$ and  $4\frac{4}{5}$ then disappear with
$T\rightarrow 0$ as a consequence of complete localization 
of the electronic states into Wigner and bubble electron solids, the
$T=0$ ground state.

A novel many-body ground state based on charge density waves in 
weak magnetic fields, and  with the uppermost Landau 
level $N\gg 1$ partially filled, has
 been proposed by Koulakov {\it et al.}\cite{Koulakov96} and independently by Moessner  and Chalker\cite{Moessner96}. Subsequently,  Fogler and Koulakov\cite{Fogler97} examined 
 the energetics of various many-electron states and showed that a transition (as
 a function of N) occurs at $\nu^{\ast}=\frac{1}{3}$ from a Laughlin to a charge  density wave  in
   Landau levels $N \geqslant 2$.  Interestingly, their analysis showed  $\nu^{\ast}=\frac{1}{5}$ 
   to be the only filling able to form a FQHE liquid in the N=2 LL. Similarly, recent
   calculations from Goerbig {\it et al.} showed  the energetics of the $\nu^{\ast}=\frac{1}{5}$
   quantum liquid to be favored in the N=2LL, and very close to a mixed phase of one- and two-electron bubbles \cite{Goerbig04}. However, given that the finite size effects (z-extent of the
  electronic wave function) in the quantum well may substantially change these energetics, 
  and that the difference in energies at  $\nu^{\ast}=\frac{1}{5}$ between Laughlin liquid 
  and charge density wave is so small, an unambiguous determination of the lowest energy
  state based on such calculations is  not possible . 
    
 Nevertheless, our results in the N=2 LL lend themselves very well to this  interpretation. 
 Importantly, the data show evidence for  a FQHE liquid {\it only} at
 filling factor $\nu^{\ast}=\frac{1}{5},\frac{4}{5}$
 in the $N=2$ LLs. Within the theoretical scenario discussed
 above, the disappearance of these FQHE states at temperatures, $T\lesssim 80$ mK,  
 would  result  from  a competition between a FQHE state and another phase present
 at the bubble and Wigner crystal phase boundary.
  
 To summarize, we have performed magnetotransport measurements in an ultra
 high-mobility GaAs/AlGaAs specimen which showed resistance minima 
 associated with re-entrant integer quantum Hall effects in the $N=2,3$ LLs.
 The data suggest
 that a magnetically driven insulator-insulator quantum phase transition may occur 
 at $T=0$. Striking new  minima are observed at 
 $\nu^{\ast}\simeq \frac{1}{5},\frac{4}{5}$ in the N=2 Landau level, but 
 only at intermediate temperatures, $80 \lesssim T \lesssim 120$ mK. 
 This may signal a possible competition between a Laughlin liquid and another phase at lower
 temperature, and support theoretical scenarios based on the
 premisses of  a charge density wave in the ground state of  $N>1$ Landau levels
 and in weak magnetic fields.

The authors would like to acknowledge helpful discussions with M.M. Fogler, B.I. Shklovskii, 
Kun Yang, Yong Chen and R.M Lewis. We are very grateful to E.C. Palm and T.P. Murphy
at the NHMFL for extensive technical assistance during the course of the experiments. 
Research funded  by the NSF under grant  \#DMR-0084173 and   \#DMR-03-52738 
and the DOE under grant \#DE-AI02-04ER46133.

\end{document}